%% file: paper.tex
\documentclass[times,twocolumn,final]{elsarticle}

\usepackage{cag}
\usepackage{framed,multirow}

\usepackage{amssymb}
\usepackage{latexsym}

\usepackage{url}
\usepackage{xcolor}
\definecolor{newcolor}{rgb}{.8,.349,.1}

\usepackage{hyperref}
\usepackage{csquotes}
\usepackage{tabularx}
\usepackage{pifont}
\usepackage{enumitem}

\usepackage[switch,pagewise]{lineno} 

\usepackage{blindtext}
\usepackage{amsmath}
\usepackage{enumitem}
\setlist[2]{noitemsep} 
\setenumerate{noitemsep} 

\newcommand*\patchAmsMathEnvironmentForLineno[1]{%
    \expandafter\let\csname old#1\expandafter\endcsname\csname #1\endcsname
    \expandafter\let\csname oldend#1\expandafter\endcsname\csname end#1\endcsname
    \renewenvironment{#1}%
        {\linenomathWithnumbers\csname old#1\endcsname}%
        {\csname oldend#1\endcsname\endlinenomath}}%
\newcommand*\patchBothAmsMathEnvironmentsForLineno[1]{%
    \patchAmsMathEnvironmentForLineno{#1}%
    \patchAmsMathEnvironmentForLineno{#1*}}%
\AtBeginDocument{%
\patchBothAmsMathEnvironmentsForLineno{equation}%
\patchBothAmsMathEnvironmentsForLineno{align}%
\patchBothAmsMathEnvironmentsForLineno{flalign}%
\patchBothAmsMathEnvironmentsForLineno{alignat}%
\patchBothAmsMathEnvironmentsForLineno{gather}%
\patchBothAmsMathEnvironmentsForLineno{multline}%
}

\journal{ISPRS Journal of Photogrammetry and Remote Sensing}

\begin{document}

\verso{Preprint}

\begin{frontmatter}

\title{Automatic reconstruction of fully volumetric 3D building models from point clouds}

\author[1]{Sebastian \snm{Ochmann}\corref{cor1}}
\author[1]{Richard \snm{Vock}}
\author[1]{Reinhard \snm{Klein}}
\cortext[cor1]{Corresponding author: 
  e-mail: ochmann@cs.uni-bonn.de}
%
\address[1]{University of Bonn, Institute of Computer Science II, Endenicher Allee 19a, 53115 Bonn, Germany}

\received{\today}

\begin{abstract}
\input{abstract}
\end{abstract}


\end{frontmatter}


\input{sections}

\bibliographystyle{cag-num-names}
\bibliography{refs}

\end{document}

%% file: abstract.tex
We present a novel method for reconstructing parametric, volumetric, multi-story building models from unstructured, unfiltered indoor point clouds by means of solving an integer linear optimization problem.
Our approach overcomes limitations of previous methods in several ways:
First, we drop assumptions about the input data such as the availability of separate scans as an initial room segmentation.
Instead, a fully automatic room segmentation and outlier removal is performed on the unstructured point clouds.
Second, restricting the solution space of our optimization approach to arrangements of volumetric wall entities representing the structure of a building enforces a consistent model of volumetric, interconnected walls fitted to the observed data instead of unconnected, paper-thin surfaces.
Third, we formulate the optimization as an integer linear programming problem which allows for an exact solution instead of the approximations achieved with most previous techniques.
Lastly, our optimization approach is designed to incorporate hard constraints which were difficult or even impossible to integrate before.
We evaluate and demonstrate the capabilities of our proposed approach on a variety of complex real-world point clouds.

%% file: sections.tex
\newcommand{\todo}[1]{\textcolor{red}{TODO: #1}}

\newcommand{\para}[1]{\paragraph{#1}}

\newcommand{\calC}{\mathcal{C}}
\newcommand{\calF}{\mathcal{F}}
\newcommand{\calR}{\mathcal{R}}
\newcommand{\calO}{\mathcal{O}}
\newcommand{\calW}{\mathcal{W}}
\newcommand{\calL}{\mathcal{L}}
\newcommand{\calRo}{\mathcal{R}_o}

\input{figures}
\input{tables}

\input{sec_introduction}
\input{sec_relatedwork}
\input{sec_overview}
\input{sec_method}
\input{sec_implementation}
\input{sec_evaluation}
\input{sec_conclusion}
\input{sec_acknowledgments}

%% file: figures.tex
\graphicspath{{images/}}

\newcommand{\figureOverview}{
    \begin{figure*}
        \begin{center}
            \includegraphics[width=\linewidth]{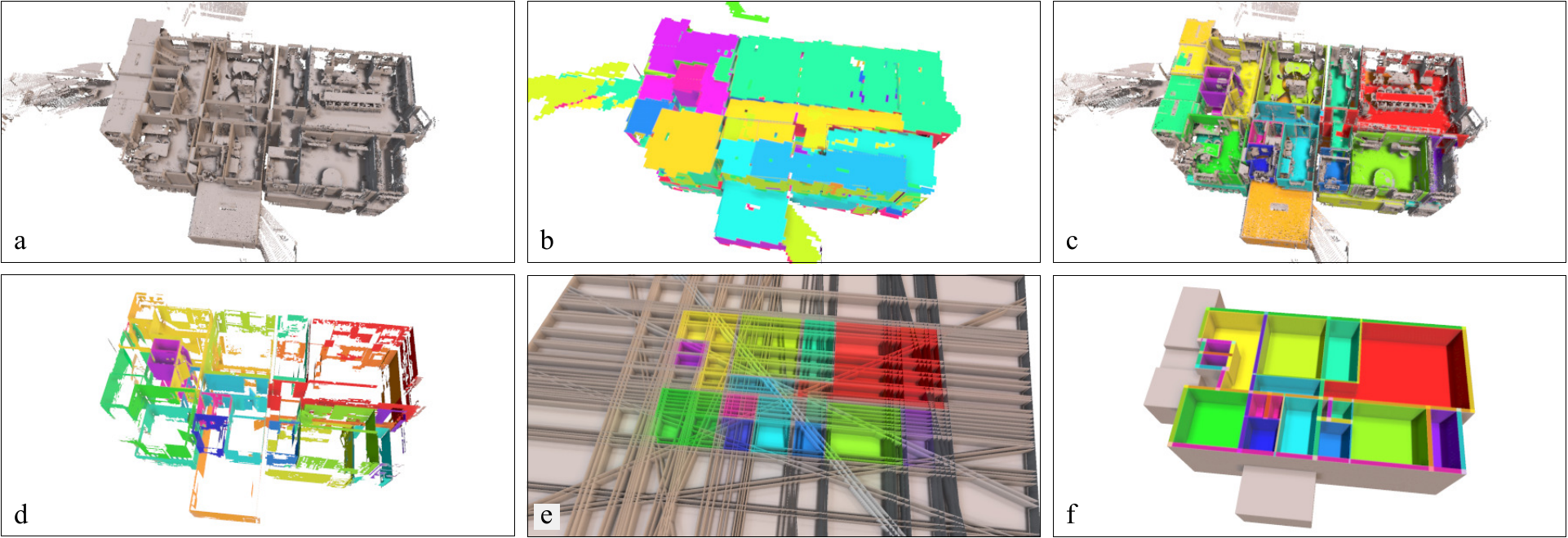}
        \end{center}
        
        \caption{
        Overview of the main steps. Ceiling of upper floor is hidden in (a), (c), and (f) for visualization purposes.
        (a) The input is a registered but otherwise unstructured and unfiltered indoor point cloud.
        (b) Planes are detected by means of RANSAC shape detection.
        (c) Outliers are automatically removed and rooms are segmented using an unsupervised clustering approach based on mutual visibility between point patches.
        (d) Detected planes are classified as horizontal slab surfaces and vertical wall surfaces (only latter shown). Surfaces are assigned multi-label support bitmaps.
        (e) A 3D plane arrangement is constructed by intersecting all planes, yielding a cell complex. Priors for rooms, outside area and surface support are estimated.
        (f) The final model consisting of interrelated room and wall volumes is obtained by solving a integer linear program in which cell labels are binary variables.
        }
        \label{fig:overview}
    \end{figure*}
}

\newcommand{\figurePlanes}{
    \begin{figure}
        \centering
        \includegraphics[width=\columnwidth]{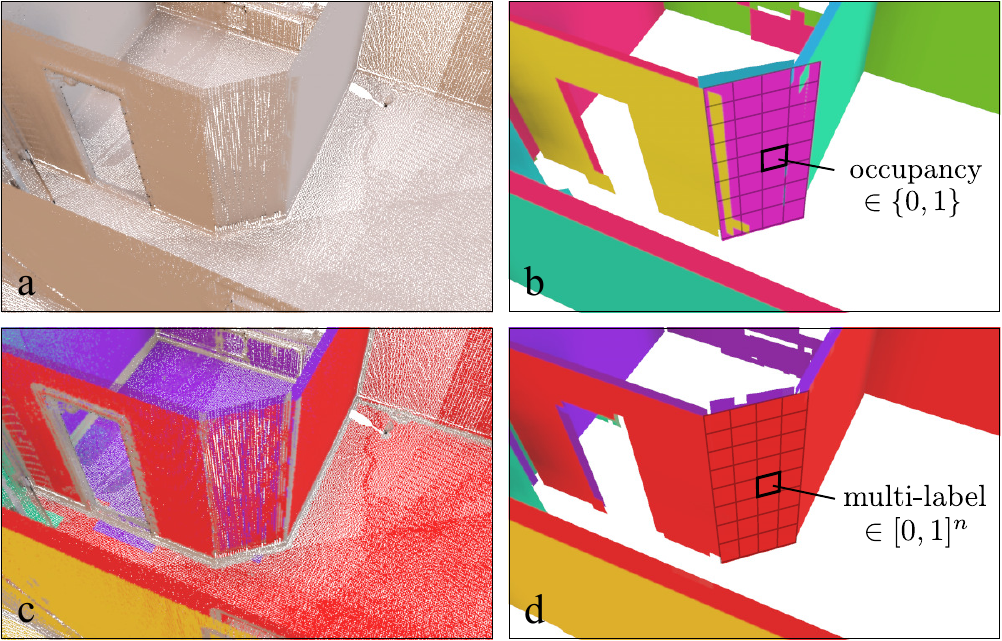}
        \caption{
        Detected planes are the basis for our reconstruction. Different kinds of bitmaps (i.e.\ grids) on the planes are used throughout the approach.
        (a) Unlabeled input point cloud.
        (b) Binary occupancy bitmaps on the detected planes (Section \ref{sec:plane_detection}) are used as a lightweight representation of support by the point cloud. Different planes are shown in different colors.
        (c) Ray casting against the occupancy bitmaps and clustering yields a segmentation of the point cloud into rooms (Section \ref{sec:point_cloud_labeling}). Different room labels are shown in different colors.
        (d) The point cloud labeling is projected into multi-label bitmaps where each pixel contains a soft assignment $[0,1]^n$ to the $n$ different room labels as defined in Section \ref{sec:point_cloud_labeling}. This is used for estimating locations of rooms in 3D space as described in Section \ref{sec:room_wall_priors}. The multi-label bitmaps are shown using the same label colors as in (c).
        In (b) and (d), only vertical planes are shown for clarity.
        }
        \label{fig:planes}
    \end{figure}
}

\newcommand{\figureDilation}{
    \begin{figure}[!b]
        \centering
        \includegraphics[width=\columnwidth]{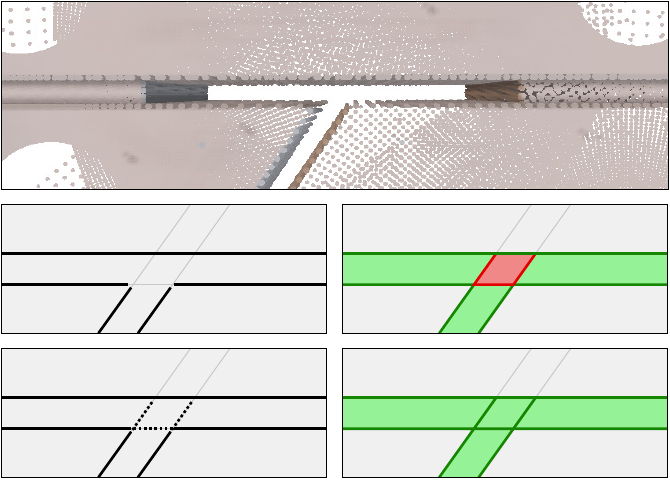}
        \caption{
        Dilation of surface support. Top: Example point cloud viewed from above. Middle: Surface support (left, thick lines) is naturally restricted to parts visible to the scanner. This leads to high costs for the wall intersection since wall surfaces are not supported by points (right). Bottom: Dilating the surface support (left, dotted lines) extends support into the interiors of walls, encouraging placement of volumetric intersections (right).
        }
        \label{fig:dilation}
    \end{figure}
}

\newcommand{\figureArr}{
    \begin{figure*}
        \includegraphics[width=\textwidth]{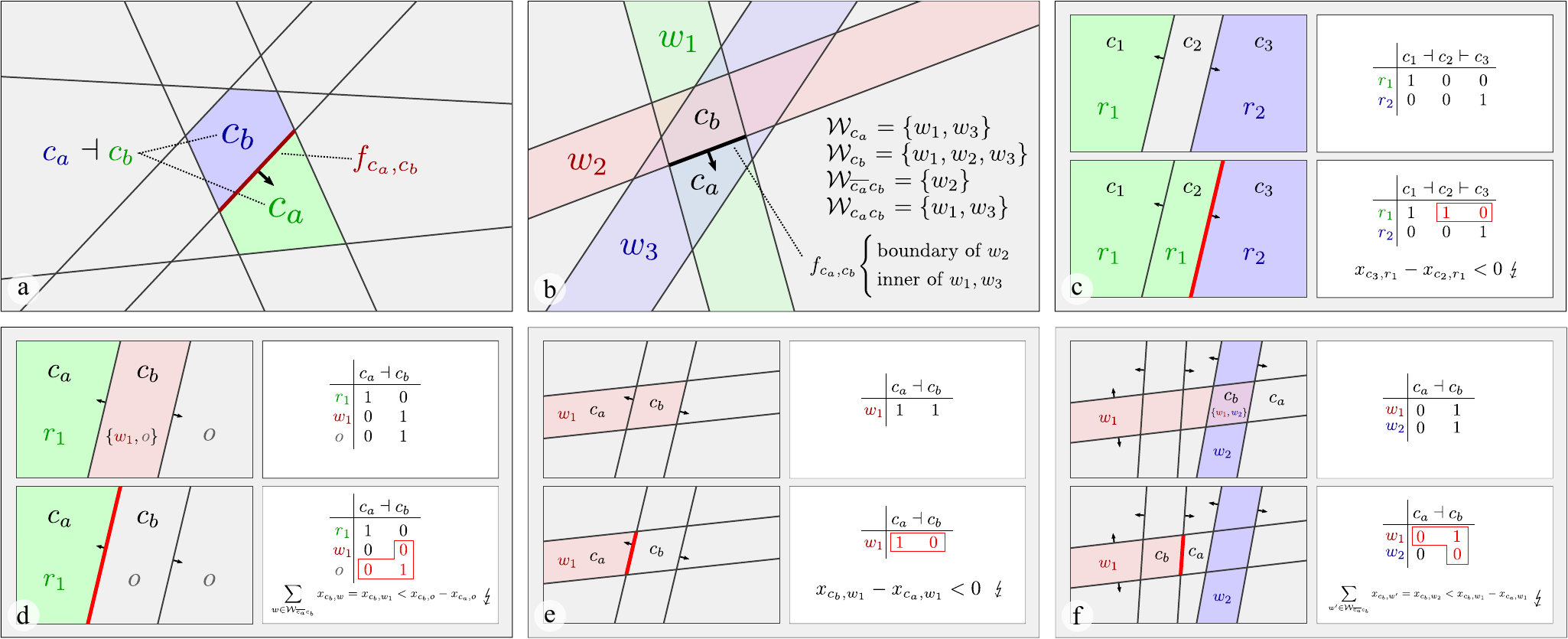}
        \caption{
        Explanation of notation and constraints.
        (a) Neighboring cells are considered as ordered pairs $c_a \dashv c_b$ with respect to normal orientation of the separating face $f_{c_a, c_b}$.
        (b) Considering a cell $c$, $\mathcal{W}_c$ is the set of wall candidates enclosing cell $c$. A face $f_{c_a,c_b}$ may be a boundary or an inner face of a set of walls. The set of boundary walls with respect to that face is $\mathcal{W}_{\overline{c_a}c_b}$, the set of inner walls is $\mathcal{W}_{c_ac_b}$.
        (c) Transitions between interior and exterior area at a face $f$ may only occur with the room label being on the positive side of $f$ (Constraint 2).
        (d) If a face $f$ is the boundary of a room, some wall needs to be active on the negative side of $f$ (Constraint 4).
        (e) A wall may end at an inner face $f$ only if the wall label is on the negative side of $f$ (Constraint 5).
        (f) If a wall $w_1$ ends at an inner face, this face must be a boundary face of at least one other active wall $w_2$. This enforces connectedness between walls.
        }
        \label{fig:arr}
    \end{figure*}
}

\newcommand{\figureComparisonOldReconstruction}{
    \begin{figure*}
        \includegraphics[width=\textwidth]{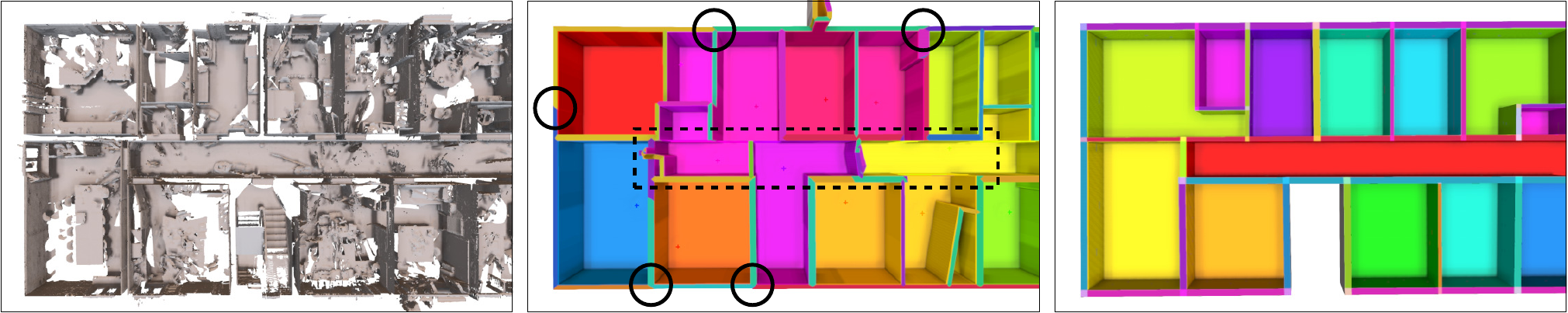}
        \caption{
        Comparison between different reconstruction approaches.
        Left: Input point cloud viewed from above.
        Center: The method described in \cite{Ochmann-2016-Parametric} may fail to regularize chains of almost coplanar walls, leading to segmented walls (circles). Also, reliance on separate scans as initial room labeling may lead to oversegmented rooms (dashed rectangle).
        Right: Our approach overcomes these issues by incorporating costs for all surfaces of volumetric wall elements, and room segmentation that is independent of scan positions.
        }
        \label{fig:comparison_old_reconstruction}
    \end{figure*}
}

\newcommand{\figureComparisonIFC}{
    \begin{figure}
        \includegraphics[width=\columnwidth]{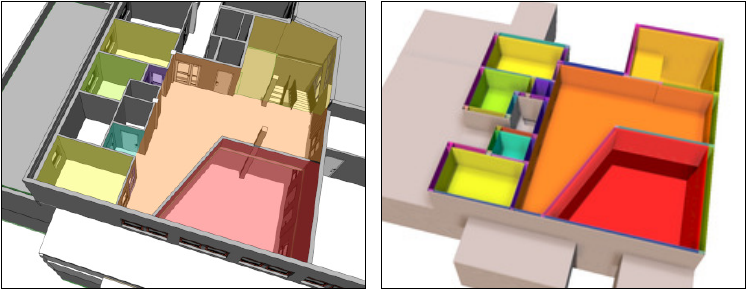}
        \caption{
        Comparison of a hand-crafted BIM model (left) and our reconstruction (right) of Dataset 5 (see Table \ref{tbl:results}). Reconstructed room labels were manually overlaid on the BIM model for reference.
        }
        \label{fig:comparison_ifc}
    \end{figure}
}

\newcommand{\figureManualConstraints}{
    \begin{figure}
        \includegraphics[width=\columnwidth]{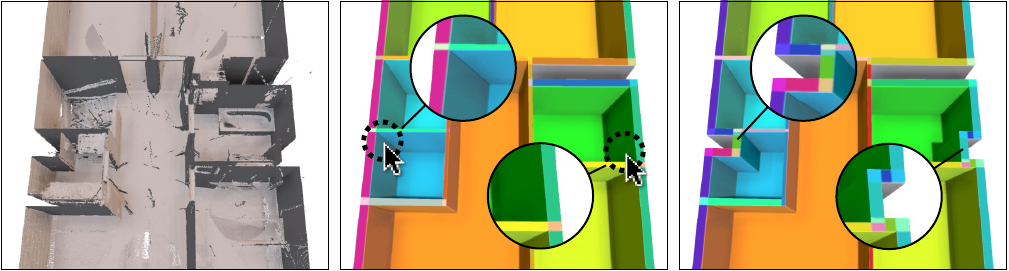}
        \caption{
        Additional constraints may be added to interactively steer the reconstruction. Small indentations of wall surfaces (left) are initially lost in the reconstruction. By forcing regions to be outside area (center), our method finds an alternative wall placement under these constraints (right).
        }
        \label{fig:manual_constraints}
    \end{figure}
}

\newcommand{\figureSynth}{
    \begin{figure}
        \includegraphics[width=\columnwidth]{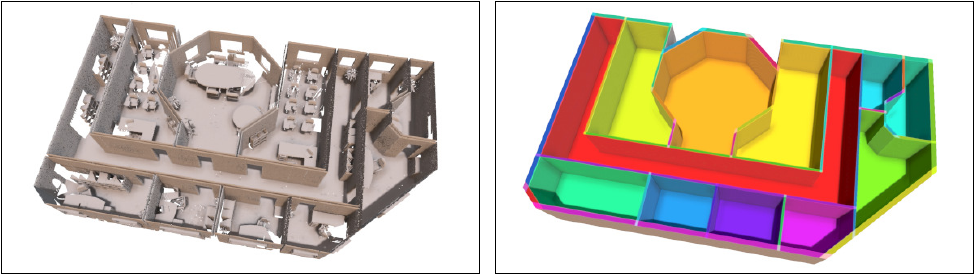}
        \caption{
        Our reconstruction result on the synthetic dataset \enquote{synth3} by the Visualization and MultiMedia Lab at University of Zurich. Rooms and walls are accurately reconstructed.
        }
        \label{fig:synth}
    \end{figure}
}

\newcommand{\figureRegularization}{
    \begin{figure*}
        \includegraphics[width=\linewidth]{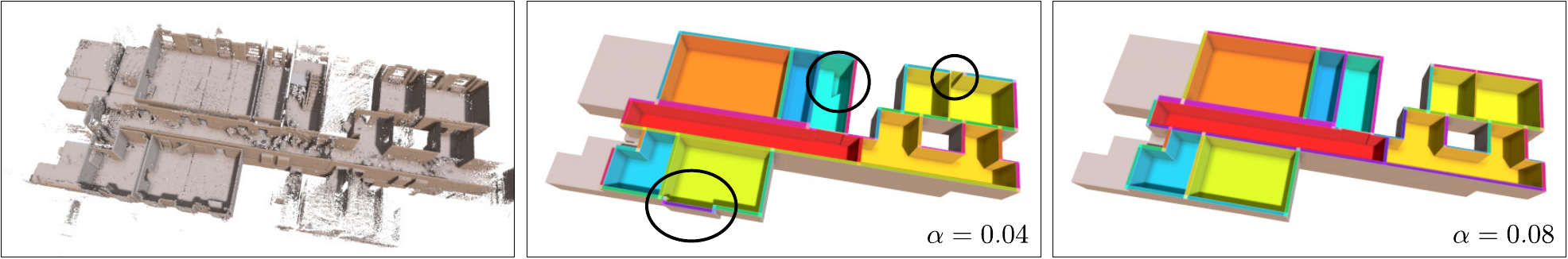}
        \caption{
        Different settings for the wall surface cost parameter $\alpha$ in Equation \ref{eq:objective} demonstrated on the dataset \enquote{Case study 2} from the ISPRS Benchmark on Indoor Modeling. Center: Our default setting of $\alpha = 0.04$ results in some walls to be fitted to windows which have high point support in this dataset. Also, a slab has a hole since floor support in staircases is often complex. Right: Increasing to $\alpha = 0.08$ leads to stronger regularization of walls and slabs.
        }
        \label{fig:regularization}
    \end{figure*}
}

\newcommand{\figureManualWall}{
    \begin{figure}
        \includegraphics[width=\columnwidth]{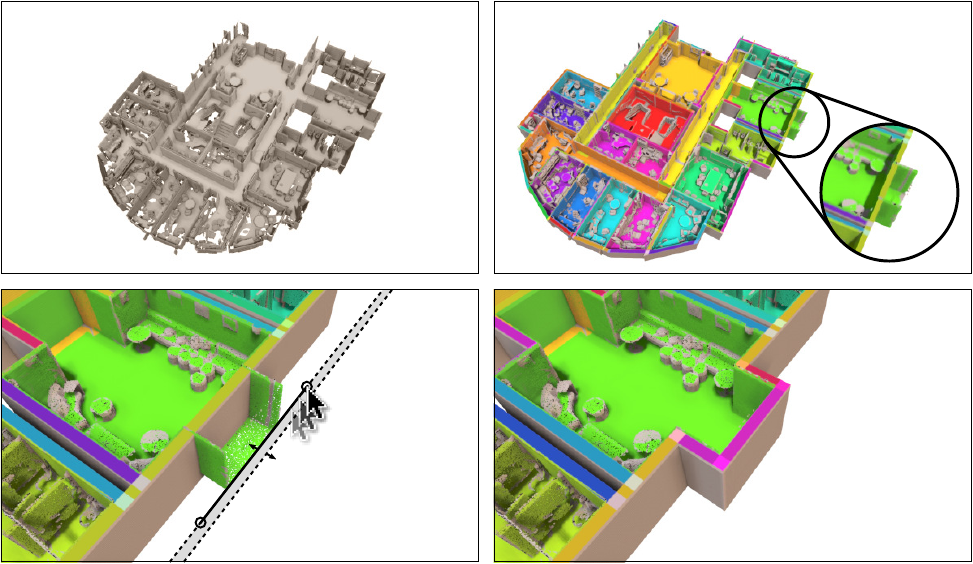}
        \caption{
        Manual addition of a wall demonstrated on the dataset \enquote{Area 3} from the Stanford 3D Large-Scale Indoor Spaces Dataset. Top right: A hallway ends without a terminating wall surface such that no wall candidate is available for enclosing the protrusion. Reconstruction and point cloud are shown overlaid. Bottom left: A wall candidate can easily be added by drawing a line, adding two opposing \enquote{virtual} wall surfaces. Bottom right: The algorithm now encloses the protrusion, using the perpendicular walls with real support in the point cloud.
        }
        \label{fig:manual_wall}
    \end{figure}
}

%% file: tables.tex
\newcommand{\tblResults}{


\begin{table*}

\small

\begin{tabularx}{\textwidth}{|l|X|X|X|}
    \hline
                                                & Dataset 1                 & Dataset 2                 & Dataset 3                     \\
    \hline                                                                         
    \textbf{Input }                             &                           &                           &                               \\
    \quad \#scans / \#points / \#pts. cleaned     & 12 / 3168600 / 2702813  & 21 / 5151388 / 4723219    & 29 / 7688111 / 5874557        \\
    \hline                                                                                                 
    \textbf{\#Entities}                         &                           &                           &                               \\
    \quad Room labels / Surfaces$^1$ / Walls$^1$& 27 / 42+5 / 37+5          & 30 / 34+7 / 28+5          & 39 / 51+5 / 39+4              \\
    \quad Cells / Variables / Constraints       & 17666 / 594748 / 1775298  & 12749 / 459373 / 1334699  & 17334 / 781794 / 2261980      \\
    \quad Nonzeros                              & 4174634                   & 3134503                   & 5315578                       \\
    \hline                                                                                                                          
    \textbf{Runtime (seconds)}                  &                           &                           &                               \\
    \quad Plane detection                       & 18.2                      & 20.1                      & 73.9                          \\
    \quad Cleaning (3 iterations)               & 14.1                      & 21.9                      & 32.7                          \\
    \quad Auto labeling                         & 6.3                       & 7.2                       & 11.4                          \\
    \quad Arrangement + Priors                  & 14.6                      & 11.5                      & 15.5                          \\
    \quad Optimization                          & 20.9                      & 7.4                       & 9.8                           \\
    \hline
    &
    \vspace{0mm}
    \includegraphics[trim={50px 0 50px 0}, clip, width=4.1cm]{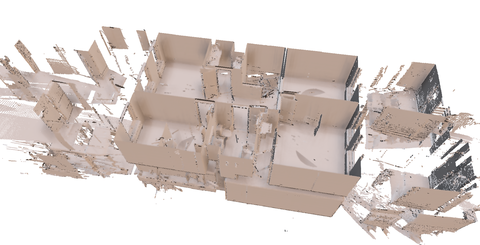}
    
    \includegraphics[trim={50px 0 50px 0}, clip, width=4.1cm]{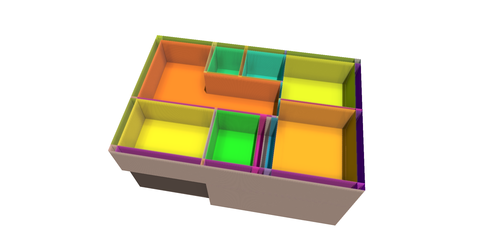}
    &
    \vspace{0mm}
    \includegraphics[trim={50px 0 50px 0}, clip, width=4.1cm]{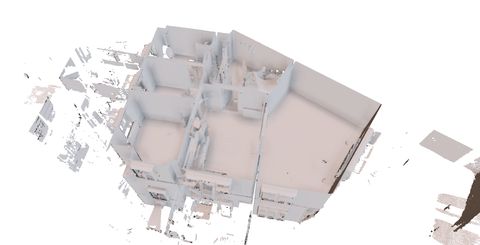}
    
    \includegraphics[trim={50px 0 50px 0}, clip, width=4.1cm]{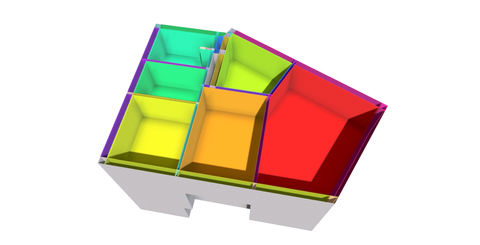}
    &
    \vspace{0mm}
    \includegraphics[trim={50px 0 50px 0}, clip, width=4.1cm]{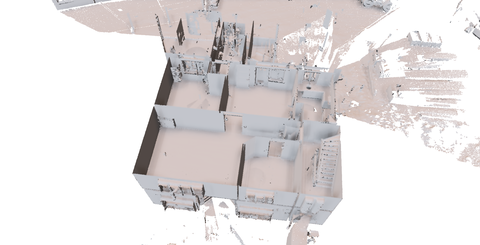}
    
    \includegraphics[trim={50px 0 50px 0}, clip, width=4.1cm]{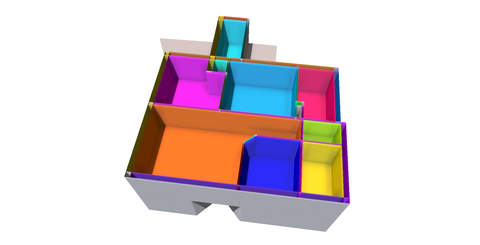}
    \\
    \hline
\end{tabularx}

\vspace{0.5cm}

\begin{tabularx}{\textwidth}{|l|X|X|X|}
    \hline
                                                & Dataset 4                 & Dataset 5                 & Dataset 6                     \\
    \hline                                                              
    \textbf{Input }                             &                           &                           &                               \\
    \quad \#scans / \#points / \#pts. cleaned   & 21 / 6452193 / 5627781    & 13 / 12409443 / 10688132  & 39 / 34964707 / 33687751      \\
    \hline                                                                                                                  
    \textbf{\#Entities}                         &                           &                           &                               \\
    \quad Room labels / Surfaces$^1$ / Walls$^1$& 49 / 61+7 / 58+5          & 29 / 48+8 / 46+7          & 108 / 72+5 / 66+5             \\
    \quad Cells / Variables / Constraints       & 42262 / 2366866 / 7035794 & 35196 / 1254848 / 3757832 & 52701 / 6041984 / 17610275    \\
    \quad Nonzeros                              & 16535885                  & 8837356                   & 41350862                      \\
    \hline                                                                                                                  
    \textbf{Runtime (seconds)}                  &                           &                           &                               \\
    \quad Plane detection                       & 51.1                      & 82.6                      & 218.6                         \\
    \quad Cleaning (3 iterations)               & 28.2                      & 59.0                      & 170.9                         \\
    \quad Auto labeling                         & 16.1                      & 56.8                      & 84.5                          \\
    \quad Arrangement + Priors                  & 38.9                      & 29.1                      & 53.6                          \\
    \quad Solving                               & 85.4                      & 42.9                      & 182.8                         \\
    \hline
    &
    \vspace{0mm}
    \includegraphics[trim={50px 0 50px 0}, clip, width=4.1cm]{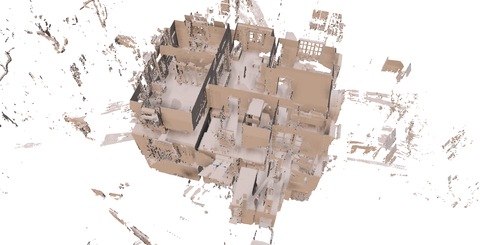}
    
    \includegraphics[trim={50px 0 50px 0}, clip, width=4.1cm]{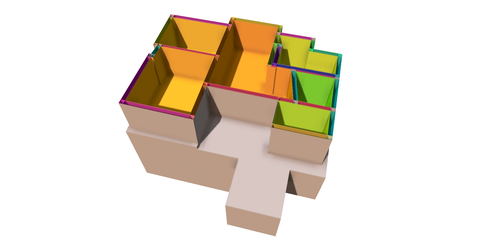}
    &
    \vspace{0mm}
    \includegraphics[trim={50px 0 50px 0}, clip, width=4.1cm]{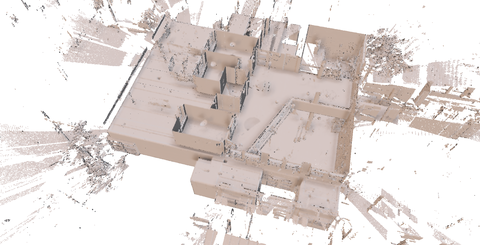}
    
    \includegraphics[trim={50px 0 50px 0}, clip, width=4.1cm]{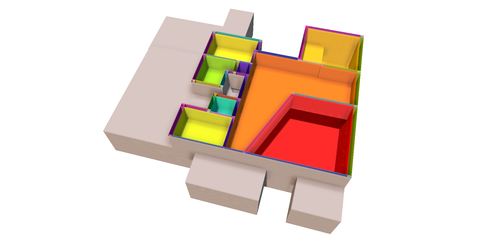}
    &
    \vspace{0mm}
    \includegraphics[trim={50px 0 50px 0}, clip, width=4.1cm]{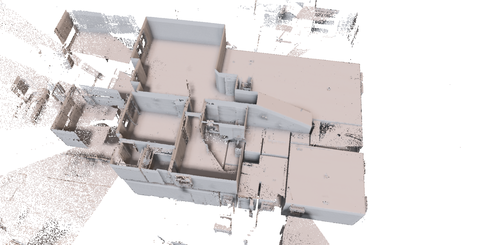}
    
    \includegraphics[trim={50px 0 50px 0}, clip, width=4.1cm]{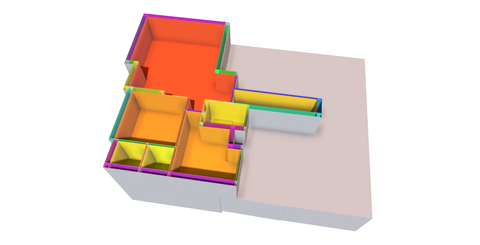}
    \\
    \hline
\end{tabularx}
\caption{
    Evaluation results on various real-world datasets. The number of scans in the point cloud is shown for reference only and is not used in our approach. The number of points is the point count after subsampling. $^1$The number of surfaces and walls is given as \emph{vertical+horizontal}.
}
\label{tbl:results}
\end{table*}
}

\newcommand{\cmark}{\text{\ding{51}}}
\newcommand{\xmark}{\text{\ding{55}}}
\newcommand{\Nmark}{\text{??}}

\newcommand{\tblApproaches}{
\begin{table*}
\footnotesize

\begin{tabularx}{\textwidth}{|l|X|X|X|X|X|X|X|}
    \hline
                                                    &   No scan positions   &   Non-Manhattan   &   Multiple rooms  &   Full 3D recons.$^1$ &   Slanted ceilings    &   Volumetric walls    \\
    \hline                                                                                                                                      
    Budroni '10 \cite{Budroni-2010-Automated3D}     &   \cmark              &   \xmark          &   \xmark          &   \xmark              &   \xmark              &   \xmark              \\
    Ad\'{a}n '11 \cite{Adan-2011-3DReconstruction}  &   \xmark              &   \cmark          &   \cmark          &   \xmark              &   \xmark              &   \xmark              \\
    Xiong '13 \cite{Xiong-2013-Automatic}           &   \xmark              &   \cmark          &   \cmark          &   \xmark              &   \xmark              &   \xmark              \\
    Mura '14 \cite{Mura-2014-AutomaticRoom}         &   \xmark              &   \cmark          &   \cmark          &   \xmark              &   \xmark              &   \xmark              \\
    Oesau '14 \cite{Oesau-2014-IndoorFeature}       &   \cmark              &   \cmark          &   \xmark          &   \xmark              &   \xmark              &   \xmark              \\
    Previtali '14 \cite{Previtali-2014-Rooms}       &   \xmark              &   \cmark          &   \xmark          &   \xmark              &   \xmark              &   \xmark              \\
    Turner '15 \cite{Turner-2015-FastAutomated}     &   \xmark              &   \cmark          &   \cmark          &   \xmark              &   \xmark              &   \xmark              \\
    Mura '16 \cite{Mura-2016-Piecewise}             &   \xmark              &   \cmark          &   \cmark          &   \cmark              &   \cmark              &   \xmark              \\
    Ochmann '16 \cite{Ochmann-2016-Parametric}      &   \xmark              &   \cmark          &   \cmark          &   \xmark              &   \xmark              &   \cmark              \\
    Ambru\c{s} '17 \cite{Ambrus-2017-Segmentation}  &   \cmark $^2$         &   \cmark          &   \cmark          &   \xmark              &   \xmark              &   \xmark              \\
    Macher '17 \cite{Macher-2017-SemiAutomatic}     &   \cmark              &   \cmark          &   \cmark          &   \xmark              &   \xmark              &   \cmark $^3$         \\
    Murali '17 \cite {Murali-2017-Scan2BIM}         &   \cmark              &   \xmark          &   \cmark          &   \xmark              &   \xmark              &   \cmark              \\
    Wang '17 \cite{Wang-2017-Modeling}              &   \xmark $^4$         &   \cmark          &   \cmark          &   \xmark              &   \xmark              &   \xmark              \\
    \textbf{Ours}                                   &   \cmark              &   \cmark          &   \cmark          &   \cmark              &   \xmark              &   \cmark              \\
    \hline
\end{tabularx}

\caption{
    Feature comparison of recent indoor reconstruction approaches.
    Notes:
    $^1$Full 3D reconstruction means that arbitrary vertical room constellations including rooms spanning multiple floors are handled correctly.
    $^2$Virtual scan positions are automatically estimated.
    $^3$Volumetric walls are generated manually in post-processing.
    $^4$Scanner trajectories are used.
}
\label{tbl:approaches}
\end{table*}
}

%% file: sec_introduction.tex
\section{Introduction}

The challenging problem of generating high-quality, three-dimensional building models from point cloud scans has been approached in a variety of ways in recent years by the computer graphics and remote sensing communities as well as in the architecture domain.
Especially for various applications in Computer Aided Design (CAD) and emerging fields such as Building Information Modeling (BIM), the reconstructed models are usually required to adhere to industry-standard specifications such as the Industry Foundation Classes (IFC).
In contrast to the representation of a building in the form of e.g.\ an unordered point cloud, a set of unconnected surfaces, or boundary meshes, a BIM/IFC model closely resembles the physical building structure by defining buildings as semantically annotated, volumetric building entities such as walls and floor slabs, usually including additional information regarding how these elements are interconnected.
%

Most previous approaches focus on the reconstruction of completely separate, planar surfaces without additional information regarding how they relate to each other \cite{Sanchez-2012-Planar}, or on representing buildings as watertight boundaries of either the whole building \cite{Oesau-2014-IndoorFeature} or separate rooms \cite{Turner-2015-FastAutomated,Mura-2016-Piecewise} and are thereby lacking in providing insights into the building structure.
Also, assumptions such as the one that stories can be globally separated by horizontal planes are very limiting in practice.
None of these approaches yields a representation which enables unhindered usage in the aforementioned scenario.
While one recent approach \cite{Ochmann-2016-Parametric} does model the measured point cloud data using volumetric building entities, the method is restricted to single-story buildings which limits its usability without laborious manual separation of the point cloud data into separate stories.
Additionally, generation of resulting wall and floor slab elements is done in a post-processing step without being integrated into the used optimization framework which may result in locally implausible results.
Other methods \cite{Murali-2017-Scan2BIM,Liu-2018-FloorNet} aiming at reconstructing true BIM models make the severe assumption that walls are positioned in a Manhattan world constellation which is often violated by real-world buildings.

Our proposed method overcomes limitations of previous approaches by alleviating the requirements on the input data and by providing a flexible optimization framework for indoor building reconstruction.
Some prior methods (e.g.\ \cite{Mura-2016-Piecewise, Ochmann-2016-Parametric}) require separate scans and scan positions to derive an initial, coarse segmentation into rooms. In contrast, our fully automatic room segmentation approach, by design, does not depend on the availability of such information and does not impose particular rules for scanning (e.g.\ one scan per room).
Furthermore, our novel integer linear programming formulation for the reconstruction problem provides flexible means to steer the reconstruction process while globally constraining the solution space to feasible solutions, thus guaranteeing a plausible model.
Additional information such as manually augmented hints may optionally be incorporated by means of hard constraints in order to further guide the reconstruction process.
%
%
While some previous approaches regularize the resulting model based on room boundary complexity, they fail to account for dependencies between surfaces related by volumetric wall elements, e.g.\ opposing surfaces between neighboring rooms.
Our formulation of the solution space based on volumetric entities enables better regularization of the model with respect to the actual volumetric walls and slabs used to represent the building.
In contrast to any previous approach, the result of our optimization immediately yields the complete geometry of all reconstructed walls and slabs as well as their volumetric intersections which allows for a direct generation of plausible BIM/IFC models.
%


In summary, the main features of our approach are:
\begin{enumerate}
    \item Fully automatic, volumetric reconstruction including volumetric intersections between elements.
    \item Flexible integration of constraints to enforce global and local properties of the resulting model.
\end{enumerate}

Our main technical contributions are:
\begin{enumerate}
    \item Automatic filtering of outliers and room segmentation of unstructured, multi-story 3D point clouds.
    \item A new formulation of the indoor reconstruction task as a linear integer programming problem that can be efficiently solved using off-the-shelf software.
\end{enumerate}

%% file: sec_relatedwork.tex
\section{Related Work}

\tblApproaches

Research on scan-to-BIM and related approaches led to a wide range of developments in recent years and still is a current topic of ongoing work. We first provide a comprehensive overview of methods dealing specifically with indoor building reconstruction which we then complement with a summary of more loosely related but complementary abstraction approaches and applications.

\subsection{Indoor building reconstruction}

The works presented in this section are closely related to our goal of indoor building reconstruction. Table \ref{tbl:approaches} summarizes and compares key features of different approaches.

Some methods aim at the generation of 2D floor plans.
Okorn et al.\ \cite{Okorn-2010-FloorPlans} model 2D floor plans by projecting detected structures into the horizontal plane and performing wall segment detection based on the Hough transform.
Ambru\c{s} et al.\ \cite{Ambrus-2017-Segmentation} reconstruct floor plans including a room labeling obtained using an energy minimization approach.
A deep neural architecture for automatic floor plan generation from RGBD video has been presented by Liu et al.\ \cite{Liu-2018-FloorNet}. Using pixel-wise predictions of floor plan geometry and semantics, integer programming \cite{Liu-2017-RasterToVector} is used to recover a vector graphics reconstruction.

Some approaches perform a reconstruction of individual rooms.
Budroni et al.\ \cite{Budroni-2010-Automated3D} reconstruct closed boundary representations of single rooms using plane sweep surface detection and a 2D line arrangement with a split-and-merge approach.
The methods by Ad\'an et al.\ \cite{Adan-2011-3DReconstruction} and Xiong et al.\ \cite{Xiong-2013-Automatic} focus on recovering detailed surface labelings, explicitly reasoning about occlusions using a ray-tracing approach.
In a similar spirit, Previtali et al.\ \cite{Previtali-2014-Rooms} perform a reconstruction of single rooms as polyhedral models including ray-tracing based reasoning about occlusions and opening detection.

Certain methods aim at the reconstruction of the building as a whole without explicitly considering room topology or segmentation.
Sanchez et al.\ \cite{Sanchez-2012-Planar} represent buildings as polygonal surface models including detection of smaller-scale structures such as parametric staircases.
Oesau et al.\ \cite{Oesau-2013-IndoorPrimitive} use a 2D cell decomposition to perform binary inside/outside labeling using a Graph-Cut based optimization.
The detail level of this approach is enhanced by Oesau et al.\ in \cite{Oesau-2014-IndoorFeature} by means of an improved line detection strategy.
With a similar goal of providing simplified environment maps for e.g.\ navigation, Xiao et al.\ \cite{Xiao-2014-Museums} employ constructive solid geometry (CSG) operations to generate a volumetric wall model. Room topology is not explicitly modeled.

Many recent methods approach the reconstruction problem in a 2.5D setting, including a segmentation into separate rooms.
Mura et al.\ \cite{Mura-2013-Robust,Mura-2014-AutomaticRoom} model buildings as 2.5D polyhedral meshes by means of constructing a 2D line arrangement and performing $k$-medoid clustering based on diffusion embeddings.
Mura et al.\ \cite{Mura-2014-ArbitraryWalls} also propose a related approach which allows arbitrary wall orientations and performs recursive clustering on a constrained Delaunay tetrahedralization.
The method by Turner et al.\ \cite{Turner-2014-FloorPlan} provides efficient means to generate 2.5D, textured meshes for e.g.\ navigation purposes including a room segmentation obtained by Graph-Cut in a triangulated environment map.
An extension providing enhanced texture mapping has been presented in \cite{Turner-2015-FastAutomated}.
The reconstruction method by Wang et al.\ \cite{Wang-2017-Modeling} models outer and inner walls by means of 2D line arrangements labeled using diffusion embeddings similar to \cite{Mura-2014-AutomaticRoom}. They also reconstruct doors using a simulated ray casting approach.
Murali et al.\ \cite{Murali-2017-Scan2BIM} present a system to quickly generate BIM models from mobile devices such as Google Project Tango, Microsoft Kinect or Microsoft HoloLens, including semantic annotations and relations between reconstructed elements. The approach is currently limited to single-story, Manhattan world buildings.

Few approaches consider the more general case of slanted walls or ceilings.
Mura et al.\ \cite{Mura-2016-Piecewise} reconstruct polyhedral room boundaries with arbitrary wall and ceiling orientations. Early rule-based classification of detected elements helps pruning invalid parts. Room segmentation is performed by clustering steered by visible surface overlap.
Mura et al.\ \cite{Mura-2017-RoomStructure} propose an extension using automatically clustered synthetic viewpoints and show applicability on complex multi-story buildings.

None of the aforementioned methods reconstruct volumetric wall and slab elements which are directly usable in a BIM setting. Few methods have approached this problem before.
Stambler et al.\ \cite{Stambler-2014-Modeling} aim to generate volumetric 3D building models using learning approaches for the classification and scoring of detected elements, and simulated annealing for optimizing the overall model. The approach makes strong assumptions about the input data, requiring both interior and exterior scans, as well as scanner positions.
Thomson et al. \cite{Thomson-2015-Geometry} generate volumetric walls from point clouds by detecting planes using a RANSAC approach and fitting suitable IFC wall entities to the detected surfaces; room volumes and topology are not explicitly modeled. They also propose a point cloud segmentation scheme based on a corresponding IFC model.
A method which explicitly represents buildings as interconnected volumetric wall elements has been presented by Ochmann et al.\ \cite{Ochmann-2016-Parametric}. They construct a 2D line arrangement of wall center lines representing pairs of opposing wall surfaces and perform a room labeling of the arrangement faces by means of a Graph-Cut based multi-label energy minimization. Multi-story buildings are not supported.
Macher et al.\ \cite{Macher-2017-SemiAutomatic} propose a semi-automatic reconstruction approach by first segmenting the input data automatically and exporting the result in an interim OBJ format, and subsequently constructing an IFC file with manual intervention in a post-processing step.

To our knowledge, our approach is the first to combine general multi-story, multi-room reconstruction with fully volumetric room and wall entities.

\subsection{Abstraction, segmentation, and reconstruction}

We now highlight some loosely related approaches which pursue more general or complementary goals which may be beneficial for tackling the reconstruction problem on different levels.
Monszpart et al.\ \cite{Monszpart-2015-RAPter} represent man-made scenes (e.g.\ buildings) by a regular arrangement of planes, taking into account non-local inter-primitive symmetry relations. Such a regularization may be useful for various arrangement-based reconstruction approaches.
A method for reconstructing lightweight, manifold, polygonal boundary models from point clouds has been presented by Nan et al.\ \cite{Nan-2017-PolyFit}. They employ an inside/outside labeling approach using binary linear programming.
Jung et al.\ \cite{Jung-2017-Morphological} generate watertight floor maps by means of skeletonization in a 2D binary occupancy map with subsequent labeling of separate rooms.
A 3D room partitioning approach using anisotropic potential fields with subsequent unsupervised clustering has been presented by Bobkov et al.\ \cite{Bobkov-2017-PotentialFields}.
Pursuing a similar goal, Ochmann et al.\ \cite{Ochmann-2014-Structural} perform a segmentation of indoor point clouds into separate rooms using a visibility-based approach. Openings between neighboring rooms are detected to obtain a room connectivity graph.
Bassier et al.\ \cite{Bassier-2018-Classification} employ a machine learning approach to classify structural elements such as walls, floors, ceilings, and beams in point cloud data.
The method by Liu et al.\ \cite{Liu-2017-RasterToVector} generates topologically and geometrically consistent floor plans from 2D raster images using an integer programming approach. While the approach is designed to work on 2D data and assumes Manhattan world geometry, the idea to enforce global properties of the resulting model using integer programming is related to our work.
Focusing on non-structural elements relevant to BIM models, Ad\'an et al.\ \cite{Adan-2018-Secondary} present an approach for detecting various important entities such as sockets, switches, signs, and safety-related items.
While the method by Son et al.\ \cite{Son-2017-Structural} does not explicitly model a building's room topology, they detect various important volumetric elements such as walls, slabs, columns and beams, also taking into account material properties and relations between elements.

\figureOverview

\subsection{Applications}

Automated scan-to-BIM methods facilitate a range of diverse applications in different areas such as construction surveillance, facility management, or energy simulations.
Garwood et al.\ \cite{Garwood-2018-Energy} propose a framework for storing building geometry in a format suitable for e.g.\ energy simulation and verification tasks, and highlight the importance of fast, automated methods for obtaining suitable models.
Hyland et al.\ \cite{Hyland-2017-Compliance} propose the usage of open standards and automatically derived BIM models from measurements for performing automated compliance control by comparing the as-built and as-designed states of buildings.
In a similar spirit, O'Keeffe et al.\ \cite{OKeeffe-2017-Validation} have developed validation approaches for determining and analyzing differences between scans and BIM models.
A prototypical approach has been presented by Brodie et al.\ \cite{Brodie-2017-BIMandScan} who propose a cloud-based platform integrating tools for generating models from and validating models against point clouds.
Krispel et al.\ \cite{Krispel-2017-Electrical} developed a method for automatic detection of power sockets and for the generation of hypotheses for electrical lines based on automatically generated building models.
An approach for integrating IFC BIM models and point cloud data in a common file format has been presented by Krijnen et al.\ \cite{Krijnen-2017-IFCSchema}. They highlight the semantically meaningful association of both worlds for documentation, structuring, annotation, synchronization and retrieval tasks.

%% file: sec_overview.tex
\section{Overview}

The input of our approach is a 3D indoor point cloud (Figure \ref{fig:overview} a) with oriented normals whose \enquote{up} direction is assumed to be the $z$-axis. If normals are not yet available, they are estimated by local Principal Component Analysis (PCA).

We first detect planes using an efficient RANSAC implementation \cite{Schnabel-2007-Primitives} (Figure \ref{fig:overview} b) and compute occupancy bitmaps for each detected plane from the respective supporting points.

The detected planes are used to automatically eliminate outlier points, and to determine point clusters corresponding to individual rooms.
This clustering is performed by means of Markov Clustering \cite{Dongen-2000-MCL} which does not require prior information about the number of rooms and results in a labeling of the point cloud (Figure \ref{fig:overview} c).

The resulting point labels are projected to the previously detected planes and discretized into multi-label bitmaps. Planes are pruned, rectified, clustered, and classified as candidates for vertical wall or horizontal slab surfaces (Figure \ref{fig:overview} d; only vertical surfaces shown for visualization purposes). Since we base our reconstruction on volumetric walls and slabs instead of single surfaces, pairs of nearby, approximately parallel surfaces are grouped to wall and slab candidates.

Based on promising previous approaches (e.g.\ \cite{Mura-2016-Piecewise,Mura-2017-RoomStructure,Ochmann-2016-Parametric,Ambrus-2017-Segmentation}), we then derive a three-dimensional \emph{arrangement of planes} from the set of wall and slab candidates (Figure \ref{fig:overview} e). To this end, all surfaces are interpreted as infinite planes and intersected with each other which results in a segmentation of 3D space into convex polyhedral cells. In particular, each wall and slab candidate is represented by a set of cells located between the respective two candidate surfaces.
Priors for the existence of different rooms and wall surfaces are estimated for each 3D cell and 2D face using the labeled surface candidates.


The main step of our approach is to find a \emph{labeling} of all cells such that each cell is either assigned to a \emph{room}, or \emph{outside space}. Additionally, volumetric \emph{walls} must be placed wherever a transition between inside and outside space takes place which is also modeled as part of the labeling problem.
The labeling should faithfully conform to the measured data and simultaneously fulfill certain constraints (e.g.\ wall connectivity) to ensure a plausible resulting model (Figure \ref{fig:overview} f).

Formulating this task as an optimization problem requires three parts:
First, we define a space of possible solutions with meaningful priors to guide the solver. The geometry of this space is given by the arrangement of planes. Priors for locations of rooms and walls in the cell complex are derived from the measured data.
Second, we need to define constraints to restrict the feasibility of a solution. They enforce that any solution satisfies predefined rules, e.g.\ a room and outside space must be separated by a wall.
Third, an objective function for assessing the quality of a solution is formulated as a cost function which is minimized under the given constraints.

After a solution is found it can easily be converted into a format suitable for rendering or exporting, e.g.\ an IFC file or a mesh, by considering the cell labeling and the boundaries between differently labeled cells.

%% file: sec_method.tex
\section{Method}
\label{sec:method}

In this Section, we provide details regarding each of the steps involved in our approach with a focus on the formulation as an optimization problem.

\subsection{Plane detection}
\label{sec:plane_detection}

Based on the widely used assumption that the coarse geometry of most buildings can be represented (or sufficiently approximated) by piecewise planar surfaces, a crucial first step of our approach is the detection of planes in the point cloud data.
To this end, an efficient RANSAC approach \cite{Schnabel-2007-Primitives} implemented in CGAL \cite{CGAL-2018-Shapes} is used.
The most important parameters are maximum point-to-plane distance, normal angle threshold, minimum number of supporting points per plane, and the probability to miss the largest plane candidate.
These can usually be chosen depending on point cloud data quality and used for a wide variety of datasets with similar characteristics (e.g.\ scanner type, density, noise level).
The supporting points of each plane are projected into occupancy bitmaps on the respective plane (Figure \ref{fig:planes} b), yielding a discretized approximation of support by measured points.
Planes with low support area (estimated using the occupancy bitmaps) are pruned later (Section \ref{sec:surface_candidates}).
Since the relatively coarse occupancy bitmaps are independent of the point cloud density, the minimum number of points for detecting a plane of the RANSAC algorithm may be set relatively low to cope with lower-resolution point clouds.

\subsection{Point cloud cleaning}
\label{sec:point_cloud_cleaning}

Real-world point clouds often contain large amounts of outlier points, often due to outside areas scanned through openings.
In order to prune outlier points early in the process, we employ a simple but very effective ray casting approach similar to \cite{Ochmann-2016-Parametric}.
From each point $p$, $n$ stochastically sampled rays $r_i$, $i = 1,\dots,n$, are cast into the hemisphere oriented into the direction of the normal at point $p$.
Ray casting is performed against the occupancy bitmaps of the previously detected primitives.
Let $h(r_i)$ be a \emph{hit} function which is $1$ if some surface was hit, and $0$ otherwise.
We approximate the probability that $p$ lies inside of the building as $in(p) = \frac{1}{n} \sum_{i=1}^n h(r_i).$
If $in(p)$ is below a given threshold (in our experiments $0.5$), $p$ is removed from the point cloud and the occupancy bitmaps of the planes are updated.
This process is iterated a small number of times.

\figurePlanes

\subsection{Point cloud labeling}
\label{sec:point_cloud_labeling}

Priors for the locations of rooms and outside area in three-dimensional space are vital for the later optimization step, even if they are coarse estimations. We formulate the estimation of priors as a point cloud labeling problem where each label represents either a room, or the outside area.

Our proposed automatic labeling approach is based on the idea that regions of the point cloud with high mutual visibility form clusters which correspond to rooms of the building.
We implement this by performing visibility tests by means of ray casting between point patches on detected surfaces which yields a visibility graph.
Nodes of this graph are then clustered by means of the Markov Clustering algorithm \cite{Dongen-2000-MCL} which determines natural clusters within the graph by flow simulation.

Point patches are constructed by generating coarse occupancy bitmaps for each plane and considering each occupied pixel as a patch with a normal identical to the respective plane normal. In our experiments, a patch size of $40 \text{cm} \times 40 \text{cm}$ was used.
We use patches instead of all points to drastically reduce the number of nodes in the visibility graph which makes the computation feasible.
Let $p_i$ be the $i$-th patch with center position $c_i$ and normal $n_i$.
For each pair $p_i$, $p_j$, $i \neq j$, ray casting between the points $c_i + \varepsilon n_i$ and $c_j + \varepsilon n_j$, with $\varepsilon := 10\text{cm}$ in our experiments, is performed.
If no surface is hit, the visibility between $p_i$, $p_j$ is set to $1$, otherwise it is set to $0$.
This yields a visibility graph whose nodes are clustered using the Markov Clustering algorithm. The computed visibility is interpreted as flow between node pairs corresponding to the respective point patches.
The main advantage of this method is that it is unsupervised and thus does not require a manual specification of the number of occurring labels.

As a result, we obtain $n$ disjoint clusters of patches which belong to different rooms and define the set of room labels $\calR := \{r_1, \dots, r_n\}$ which will be used throughout the remainder of the reconstruction process. Each point of the point cloud is assigned the room label of the respective point patch.
Note that the number of room labels $n$ may be larger than the number of rooms that will actually be contained in the final reconstruction.

\subsection{Surface candidates}
\label{sec:surface_candidates}

\figureDilation

The detected planes usually include many surfaces which are not part of walls, floors and ceilings.
Even correctly detected surfaces will generally not be perfectly vertical or horizontal.
We thus apply a pruning, classification and rectification step to extract two sets of candidates for wall and slab surfaces.
The occupancy bitmaps are used to estimate the support area of each surface independently of point cloud density.
Planes with support below an area threshold as well as planes which are not approximately vertical or horizontal are discarded.
The remaining planes are classified as wall or slab surface candidates depending on their normal direction, and adjusted to be perfectly horizontal or vertical.

As a prerequisite for later room prior estimation (Section \ref{sec:room_wall_priors}), each surface is also assigned a multi-label support bitmap with continuous values in $[0, 1]$ for each room label in $\calR$ (Figure \ref{fig:planes} d).
This provides a soft-assignment of different regions of each surface to different room labels.
The label bitmaps are generated by projecting all supporting points onto the respective surface and averaging the previously determined point labels within each pixel.

Furthermore, we \emph{dilate} the support bitmaps.
The rationale is that reconstructed walls with no surface support by the point cloud data are penalized by a cost function defined later in Section \ref{sec:optimization}.
Since we reconstruct wall intersections volumetrically, placing the respective wall entities in between rooms would cause high costs since surface support is naturally restricted to regions that are visible to the scanner (Figure \ref{fig:dilation}, middle row).
By slightly extending the surface support, we encourage construction of intersecting wall entities in regions with nearby surface support (Figure \ref{fig:dilation}, bottom row).


\subsection{Wall and slab candidates}
\label{sec:wall_slab_candidates}

Since our approach is based on the notion of \emph{volumetric} walls and slabs instead of single surfaces, the next step is to determine pairs of opposing surfaces forming potential building elements.
To this end, a simple pairing procedure is employed.
For each surface, we search a matching, approximately parallel surface with opposing normal orientation within a user-defined distance and angle threshold.
If a match is found, the two surfaces are paired to form a wall or slab candidate.
It should be noted that a single surface may thus be part of multiple pairs.
%
For surfaces without any matching counterpart, \emph{virtual} surfaces with a user-defined distance are added to the set of surfaces.
This is usually the case for outside walls for which only the inner side has been scanned.
This augmentation is important since interior spaces are required to be bounded by volumetric walls or slabs.
The $m$ generated candidates constitute the set of wall/slab labels $\calW := \{w_1, \dots, w_m\}$.
Which of these walls and slabs are contained in the final model is decided by the optimization described in Section \ref{sec:optimization}.

\subsection{Arrangement of planes}
\label{sec:arrangement}

The geometry of the search space for finding an optimal constellation of rooms, walls and slabs is modeled as an \emph{arrangement of planes} and the \emph{3D cell complex} induced thereby. It is constructed by intersecting all (infinite) planes of the wall and slab candidate surfaces with each other.
Since vertical walls and horizontal slabs are treated identically, we will hereafter refer to both simply as \emph{walls}.

Cells of the arrangement are convex, three-dimensional subsets of the space inside and outside of the building.
Each cell belongs either to a room, or the outside area.
Additionally, walls may be placed in cells that are part of the outside area.
Constraints such as that a cell may belong to at most \emph{one} room, or that walls may \emph{only} occur in the outside area (e.g.\ between rooms) are formulated as constraints for the optimization problem in Section \ref{sec:optimization}.

Faces between neighboring cells are convex, two-dimensional subsets of regions on the planes of wall surfaces. Each face may separate different regions (e.g.\ a room and a wall) from each other.

\figureArr

\subsection{Volume and surface priors}
\label{sec:room_wall_priors}

For guiding the optimization, two kinds of priors are estimated from the data.
First, volumetric priors for the existence of different rooms as well as outside area are estimated for each 3D cell of the arrangement.
Second, support by the point cloud data is estimated for each 2D face between neighboring cells.

\para{Preparations}
The arrangement consists of cells $\calC := \{c_1, \dots, c_p\}$.
For two cells, the notation $c_a \dashv c_b$ means that $c_a, c_b$ are neighboring and the normal of the separating oriented face $f_{c_a,c_b}$ points towards $c_a$ (Figure \ref{fig:arr} a).
The set of all oriented faces is denoted as $\calF := \{f_{c_a,c_b} \ \vert \ c_a \dashv c_b\}$.
For brevity, we write $f$ instead of $f_{c_a,c_b}$ if the specific incident cells are irrelevant.

The set of room labels is $\calR := \{r_1, \dots, r_n\}$ with $n$ being the number of room clusters as introduced in Section \ref{sec:point_cloud_labeling}, and the set of wall labels is $\calW := \{w_1, \dots, w_m\}$ with $m$ being the number of generated wall candidates as introduced in Section \ref{sec:wall_slab_candidates}. We furthermore define an additional \emph{outside} label $\calO := \{o\}$.
As detailed later, the outside label is used for cells that are not the interior space of a room.
The union of rooms and outside labels is denoted $\calRo := \calR \cup \calO$; the set of all labels is $\calL := \calR \cup \calO \cup \calW$.

Let $\mathcal{C}_{w} \subseteq \mathcal{C}$ be the set of cells that are \emph{contained} in wall candidate $w$, i.e.\ all cells that are located between the two surfaces of $w$. Conversely, $\calW_c := \left\{ w \in \mathcal{W} \ \vert \ c \in \mathcal{C}_{w} \right\}$ is the set of walls that \emph{contain} cell $c$.

For a particular cell pair $c_a, c_b$, we define the set $\mathcal{W}_{\overline{c_a}c_b}$ of walls that are contained in cell $c_b$ but \emph{not} in $c_a$, i.e.
$$ \mathcal{W}_{\overline{c_a}c_b} := \mathcal{W}_{c_b} \setminus \mathcal{W}_{c_a}. $$
The separating face $f_{c_a, c_b}$ is called a \emph{boundary} face of the walls in $\calW_{\overline{c_a}c_b}$.
Analogously, we define the set $\mathcal{W}_{c_ac_b}$ of walls that are contained in \emph{both} $c_a$ and $c_b$, i.e.
$$ \mathcal{W}_{c_ac_b} := \mathcal{W}_{c_a} \cap \mathcal{W}_{c_b}. $$
The separating face $f_{c_a, c_b}$ is called an \emph{inner} face of the walls in $\calW_{c_ac_b}$.
These definitions are exemplified in Figure \ref{fig:arr} b.

It should be noted that an inner face of a wall $w$ is always the boundary face of another wall (which is often approximately perpendicular to $w$). As an example, in Figure \ref{fig:arr} b, face $f_{c_a,c_b}$ is an inner face of wall $w_1$ and a boundary face of the intersecting wall $w_2$. This will become important for the definition of the optimization constraints in Section \ref{sec:optimization}.

\para{Room and outside priors}
To estimate probabilities where different rooms and outside area are located in 3D space, we estimate a volumetric prior function $p_\calC(c, l) : \calC \times \calRo \to [0, 1]$ which returns a high value iff a label $l$ is likely to occur within a cell $c$.
To this end, we perform stochastic ray casting from sampled points in 3D space and average previously computed room labels on surfaces visible from each point.
For each cell $c$, $k$ random points are sampled within $c$. To draw enough samples for narrow cells, which are very common due to parallel surfaces, $k$ is chosen proportional to
$$\max(volume(c), diameter(c)).$$
Centered at each sampled point, $d$ rays are cast into random directions.
$p_\calC(c, l_i), i = 1, \dots, n,$ is then set to the average over all observed room labels. Rays hitting the back side of surfaces, as well as rays without surface intersections, are counted as \emph{outside}.

\para{Face support priors}
In addition to the volumetric room and outside prior function, we estimate a face support function $p_\calF(f) : \calF \to [0, 1]$ which returns a high value iff a face $f$ is supported by the point cloud.
This function is later used for selecting probable wall candidates and regularizing the optimization result.
To estimate $p_\calF(f)$ for a face $f$, we first sample $k$ random points within $f$ where $k$ is proportional to
$$\max(area(f), diameter(f)).$$
Subsequently, all sampled points are projected onto the surface from which face $f$ was generated in the arrangement.
$p_\calF(f)$ is then set to the ratio between the number of sampled points lying within the support approximated by the occupancy bitmap of the respective surface to the total number of sampled points.

\subsection{Cell complex optimization}
\label{sec:optimization}

For finding an optimal labeling of all cells, we employ a 0-1 integer linear programming approach in which binary variables for each cell are interpreted as room, outside, and wall label assignments to cells.
This approach has the advantage that a set of rules to be fulfilled by any feasible solution can be formulated as hard constraints.
Approximate multi-label methods based on e.g.\ Graph Cuts \cite{Boykov-2001-Fast} are more restricted regarding the family of objective functions and constraints that can be used and may fail to find good solutions if the objective is not sufficiently smooth.
We first discuss the set of constraints imposed on our model before defining the objective function.

\para{Preparations}
Each of the binary variables $$x_{c,l} \in \{0, 1\},\ c \in \calC,\ l \in \calL,$$ of our optimization is a binary assignment of a label $l$ to a cell $c$. A value of $1$ means that the label is \emph{assigned} or \emph{active}.
It should be noted that a cell is not necessarily assigned only a single label. In particular, a cell can be assigned the outside label \emph{and} a nonempty set of wall labels at the same time as defined by the constraints below. Also, cells where walls intersect are assigned \emph{all} labels of the intersecting walls.
We also use the notion of \emph{inner} and \emph{boundary} faces as defined in Section \ref{sec:room_wall_priors}.

\textit{Constraint 1.} Each cell $c$ must be assigned exactly one label from $\calRo$, i.e.
\begin{equation}
    \forall c \in \calC: \ \sum_{r \in \calRo} x_{c,r} = 1.
\end{equation}

\textit{Constraint 2.} At boundary faces of room interiors, the room label may only occur on the \emph{positive} side of the separating face, i.e.
\begin{equation}
    \forall f_{c_a,c_b} \in \calF \ \forall r \in \calR: \ x_{c_a,r} - x_{c_b,r} \geq 0,
\end{equation}
as shown in Figure \ref{fig:arr} c. Note that this constraint implies that two different room labels $r_p \neq r_q$ cannot be directly neighboring since this would violate the constraint for one of the room labels. As a consequence, this avoids \enquote{paper thin} walls between rooms since they must be separated by outside area, thereby following the physical nature inherent to walls.

\textit{Constraint 3.} Wall labels may only occur in cells which are assigned the outside label, i.e.
\begin{equation}
    \forall c \in \calC \ \forall w \in \calW_c: \ x_{c,w} \leq x_{c,o}.
\end{equation}

\textit{Constraint 4.} The boundary faces of room interiors must also be the boundary faces of an active wall, i.e.
\begin{equation}
    \forall f_{c_a,c_b} \in \calF: \ \sum_{w \in \calW_{\overline{c_a}c_b}} x_{c_b, w} \geq x_{c_b, o} - x_{c_a, o},
\end{equation}
as illustrated Figure \ref{fig:arr} d. This constraint implies that there cannot be a transition between room interior and outside area without activating a wall at all faces where the transition occurs.

\textit{Constraint 5.} At wall boundaries which occur at inner faces, the wall label must be on the negative side of the respective faces, i.e.
\begin{equation}
    \forall f_{c_a,c_b} \in \calF \ \forall w \in \calW_{c_ac_b}: \ x_{c_b,w} - x_{c_a,w} \geq 0,
\end{equation}
as exemplified Figure \ref{fig:arr} e. This constraint is a prerequisite for Constraint 6 as well as the objective function which require the left-hand side expression to be nonnegative.

\textit{Constraint 6.} A wall may end at an inner face only if this face is a boundary face of at least one other active wall, i.e.
\begin{equation}
    \forall f_{c_a,c_b} \in \calF \ \forall w \in \calW_{c_ac_b}: \ \sum_{w' \in \calW_{\overline{c_a}c_b}} x_{c_b, w'} \geq x_{c_b,w} - x_{c_a,w},
\end{equation}
as depicted Figure \ref{fig:arr} f. This constraint enforces that walls are interconnected at their endpoints since it disallows that a wall ends at an inner face without it coinciding with a boundary face of an active wall.

\para{Objective function}
To determine the optimal labeling, we define a cost function $F_C$ for a solution over the for cell complex $C$ of the form
\begin{equation}
    \label{eq:objective}
    F_C := -R_\calC + \alpha(W_{\calF_b} + W_{\calF_i}),
\end{equation}
consisting of the following terms.
The volumetric room and outside area fitness term $R_\calC$ rewards the assignment of the most likely labels for each cell $c \in \calC$ and is defined as
\begin{equation}
    \label{eq:RC}
    R_\calC := \sum_{c \in \calC} \sum_{r \in \calRo} x_{c,r} \cdot p_\calC(c,r) \cdot volume(c),
\end{equation}
where $x_{c,r}$ denotes the binary variable for the assignment of label $r$ to cell $c$ and $p_\calC(c,r)$ represents the volumetric room and outside prior (Section \ref{sec:room_wall_priors}), weighted by the volume of cell $c$.
Note that this term is included with a negative sign within $F_C$ such that its value is being maximized.
The wall face cost terms $W_{\calF_b}$ and $W_{\calF_i}$ penalize placement of walls in terms of the required boundary and inner face areas, respectively. This penalty is attenuated for faces with high support. The terms are defined as
\begin{equation}
    \label{eq:WFb}
    W_{\calF_b} := \sum_{f_{c_a,c_b} \in \calF} \sum_{w \in \calW_{\overline{c_a}c_b}} x_{c_b,w} \cdot (1-p_\calF(f_{c_a,c_b})) \cdot area(f_{c_a,c_b}),
\end{equation}
and
\begin{equation}
    \label{eq:WFi}
    W_{\calF_i} := \sum_{f_{c_a,c_b} \in \calF} \sum_{w \in \calW_{c_ac_b}} (x_{c_b,w} - x_{c_a,w}) \cdot (1-p_\calF(f_{c_a,c_b})) \cdot area(f_{c_a,c_b}),
\end{equation}
where $x_{c_a,w}, x_{c_b,w}$ are the binary variables for the assignment of the wall label $w$ to the cells $c_a, c_b$ respectively, $p_\calF(f_{c_a,c_b})$ is the face support prior (Section \ref{sec:room_wall_priors}), and $area(f_{c_a,c_b})$ is the area of face $f_{c_a,c_b}$.
It should be noted that $(x_{c_b,w} - x_{c_a,w}) \in \{0, 1\}$ due to Constraint 5.
Also note that in Equation \ref{eq:WFb}, it suffices to consider $x_{c_b,w}$ since for a boundary face of wall $w$, $x_{c_a,w}$ does not exist (i.e.\ $x_{c_a,w}$ can be considered to be zero).

We then minimize $F_C$ s.t.\ Constraints 1-6 using the Gurobi Optimizer \cite{gurobi}.

Note that in our experiments, we added the following constraint which gave a small performance improvement although it is already implied by Constraints 1-2. At boundary faces of outside area, the outside label may only occur on the \emph{negative} side of the separating face, i.e.
\begin{equation}
    \forall f_{c_a,c_b} \in \calF: \ x_{c_a, o} - x_{c_b, o} \leq 0.
\end{equation}
We attribute this slight performance improvement to heuristics used by the particular optimizer implementation.

\subsection{Optimization result}

The result is an assignment of each cell to either one room, or the outside area. Cells which are assigned the outside area may also be assigned a nonempty set of walls.
On the one hand this provides a dense segmentation of space into rooms and outside space. Volumes to which multiple walls are assigned are (volumetric) intersections of the respective walls. Since the underlying data structure provides adjacency information between all cells, semantic information like room adjacency and wall incidence is immediately available, e.g.\ for navigation or simulation purposes.
On the other hand this information is closely related to the definition of building elements in BIM formats like IFC. This enables immediate transfer of the results into standard architecture software and integration into existing BIM pipelines.

%% file: sec_implementation.tex
\section{Implementation details}

Input point clouds were subsampled to a minimum point distance of 2 cm. Plane detection was performed using a plane distance threshold of 1 cm, a point cluster epsilon of 20 cm, a normal threshold of about $6^\circ$ ($18^\circ$ for the \enquote{Case study 2} dataset), minimum support of 1000 points and miss probability of 0.001.
Multi-label bitmaps had a resolution (pixel size) of 10 cm, occupancy bitmaps had a resolution of 20 cm.
Three ray casting iterations were performed for point cloud cleaning.
For automatic labeling, MCL was used with default parameters (inflation set to 2.0) in multi-threaded mode.
The surface cost weight $\alpha$ in Equation \ref{eq:objective} was empirically chosen as 0.04. We used PCL 1.8.1 \cite{PCL}, CGAL 4.12 \cite{CGAL-2018-Shapes,CGAL-2018-Arrangements}, MCL 14-137 \cite{Dongen-2000-MCL}, Gurobi 8.0.1 \cite{gurobi}, and NVIDIA OptiX 5.0 for GPU-based ray casting under Linux on a 6-core Intel i7 CPU and a NVIDIA GeForce GTX 980 GPU.

%
%

%% file: sec_evaluation.tex
\tblResults

\section{Evaluation}

We evaluate the reconstruction quality and performance of our approach on a variety of datasets and show comparisons with groundtruth IFC and related work. Furthermore, we exemplify the flexibility of our integer linear programming approach by specifying additional constraints to modify and guide the resulting reconstruction in an intuitive manner.

\para{Datasets}
We used a variety of real-world datasets and one synthetic dataset for our evaluation. 
Table \ref{tbl:results} shows six multi-story point clouds measured using terrestrial laser scanners.
These datasets were provided by The Royal Danish Academy of Fine Arts
Schools of Architecture, Design and Conservation (CITA).
The Table lists properties of the input data including the number of points and scans, as well as quantities derived during reconstruction such as the number of room labels, extracted surfaces, wall candidates, etc.
It also shows runtime measurements of the main processing steps.
We also tested our approach on publicly available datasets provided by other research groups.
Figure \ref{fig:synth} shows the dataset \enquote{synth3} by the Visualization and MultiMedia Lab at University of Zurich, Figure \ref{fig:regularization} depicts the dataset \enquote{Case study 2} from the ISPRS Benchmark on Indoor Modeling \cite{ISPRS}, and Figure \ref{fig:manual_wall} shows the dataset \enquote{Area 3} from the Stanford 3D Large-Scale Indoor Spaces Dataset \cite{Buildingparser}.
We used the latter two for demonstrating different parameters and interactive modification as described below.

\figureSynth

\figureComparisonIFC

\figureComparisonOldReconstruction

\figureManualConstraints

\para{Reconstruction quality}
Our reconstruction approach generally worked well on the test datasets without any dataset-specific tuning.
Automatic outlier removal reliably ignored even large-scale clutter scanned through windows in e.g.\ Datasets 1, 3, and 6.
In some cases, particularly thick walls (bottom region of Dataset 1, top region of Dataset 2) were reconstructed as two thinner, parallel wall elements which may be a matter of interpretation.
Increasing the maximum thickness of generated wall candidates in these cases can help recognizing such cases as single walls.
A few cases of room-oversegmentation can be observed.
In Dataset 4, the large central room is split into a larger L-shaped part (orange) and a smaller room (green, to the right of the building) without a real wall separating the reconstructed rooms in the point cloud data.
In Dataset 5, indentations of the central room (orange) were reconstructed as small, separate rooms (cyan, purple).
Since our approach currently only considers horizontal ceilings, the slanted ceiling of the staircase in Dataset 6 (yellow, elongated room) is reconstructed as a horizontal structure (see also Limitations below).

\para{Runtime}
Total runtime for the reconstruction of the test datasets lies in the range of one minute (Datasets 1, 2) to 10 minutes (Dataset 6).
The runtime of primitive detection is mainly dependent on the CGAL implementation, and the time for solving the optimization problem is the runtime of the Gurobi optimizer. The runtime for auto labeling contains the time for our raycasting and clustering using the Markov Cluster Algorithm.
Runtime of the optimization mainly depends on the complexity of the plane arrangement, which in turn depends on the number of detected surfaces since every surface introduces global splits in the cell complex. Therefore a tradeoff between reconstructing details (i.e.\ small surfaces) and computational feasibility must be made. In our experiments, we thus chose a minimum estimated area of 2 m$^2$ for vertical surfaces and 5 m$^2$ for horizontal surfaces.

\para{Comparison to IFC}
For Dataset 5 a corresponding, professionally made BIM model in IFC format was available. Figure \ref{fig:comparison_ifc} shows a comparison between our reconstruction and the BIM model. Colors of the reconstructed rooms were manually overlaid on the IFC model on the left-hand side. All rooms that were part of the scans mostly match the groundtruth BIM. The upper story of the building is connected to the lower story through a large horizontal opening. These areas were reconstructed as two separate rooms (red and orange) and the railing at the edge of the gallery was reconstructed as walls. The small cyan and purple rooms are an oversegmentation of the upper floor, probably due to the dilated surface support. However, this error can easily be fixed manually.

\figureManualWall

\figureRegularization

\para{Comparison to related work}
A comparison between reconstructions by the approach described in \cite{Ochmann-2016-Parametric} and our method is shown in Figure \ref{fig:comparison_old_reconstruction}. In addition to fundamental advantages of our approach such as reconstruction of multiple stories, two crucial differences are particularly notable.
First, our approach results in stronger regularization of wall elements where using multiple different, similar walls to represent the building would be unnecessary. The approach in \cite{Ochmann-2016-Parametric} leads to jumps between different, almost coplanar walls (Figure \ref{fig:comparison_old_reconstruction}, center, black circles) instead of using longer, continuous walls. This can be explained by the principle that the approach separated rooms by wall center lines in 2D such that jumping from one wall to an almost coplanar wall resulted in almost no penalty in the cost function. In our case, a fully volumetric wall element would need to be added to the model to connect the parallel walls, resulting in relatively high costs.
Second, the approach in \cite{Ochmann-2016-Parametric} relies on given, separate scans and their positions for estimating an initial room segmentation. This leads to an oversegmentation of the hallway (Figure \ref{fig:comparison_old_reconstruction}, center, dashed rectangle) since it tries to reconstruct one room per scan. Our method works independently of separate scans and estimates a room segmentation by unsupervised clustering.

\para{Interactive modification}
Our linear programming approach allows for additional constraints to be easily added.
One example for manual post-processing of the reconstruction results by interactively adding hard constraints is shown in Figure \ref{fig:manual_constraints}. In this case the indentation of the wall surfaces on the left and right sides of the building were lost by the regularization of the model as can be seen in Figure \ref{fig:manual_constraints}, center. The user has the option to add constraints such as forcing inside area, outside area, wall, no wall, etc.\ by clicking at the desired location. In this case, the highlighted locations were forced to be outside area. The algorithm then finds the next best option, placing new walls that fulfill all constraints as shown in Figure \ref{fig:manual_constraints}, right.
Another example is shown in Figure \ref{fig:manual_wall} where a hallway ends without any terminating wall surface in the input data. Since the algorithm has no wall candidate available, it cannot enclose the protruding room area. By adding a virtual wall candidate by means of simply drawing a line, the algorithm is able to include the protrusion in the reconstructed model, automatically using the perpendicular wall surfaces that are present in the input data.
Different choices of the wall surface penalty weight $\alpha$ in Equation \ref{eq:objective} control global regularization strength. Figure \ref{fig:regularization}, center shows a reconstruction where some wall and slab elements are slightly misplaced due to relatively strong surface support at windows. Increasing $\alpha$ from our default of 0.04 to 0.08 leads to stronger regularization as shown in Figure \ref{fig:regularization}, right.

\para{Limitations}
One technical limitation of our current implementation is that slanted walls, floors or ceilings are not taken into account although this is not an inherent limitation of our approach.
The reason for our decision not to include these elements is that the construction of the 3D cell complex needs to be exact to guarantee the integrity of the data structure (e.g.\ cell neighborhood).
Unfortunately, computing the cell complex in 3D induces numerical problems and currently we do not have a stable implementation for this task at our disposal.
We thus opted to use the numerically stable implementation of 2D arrangements in CGAL \cite{CGAL-2018-Arrangements} and extend it to 3D by stacking 2D arrangements separated by horizontal planes.
A numerically stable extension of arrangements supporting general slanted planes would be an interesting direction for future research which we consider to be outside the scope of this paper.
Processing of very large datasets may also require further optimizations to make them computationally feasible. In particular, using a global plane arrangement results in a large increase of cells and thus variables in the optimization model with every additional detected surface. More sophisticated selection of potential surfaces, and improved optimization methods, e.g.\ splitting the problem into smaller subproblems, are targets for further research.
Last but not least, our current algorithm is not able to identify and include important architectural structures overarching the whole building like the pillars that are included in the hand-crafted model in Figure \ref{fig:comparison_ifc}. Automatically identifying such structural elements and incorporating them into the automatic reconstruction is also an interesting direction for future research.


%% file: sec_conclusion.tex
\section{Conclusion and future work}

We have presented a novel approach to tackle the indoor building reconstruction problem from point clouds using integer linear programming. In contrast to previous methods, our approach reconstructs fully volumetric, interconnected wall entities and room topology on multi-story buildings with weak assumptions on the input data. The resulting models are very close to the requirements needed for Building Information Modeling tasks including volumetric representations of room spaces and wall entities, and their interrelations. Additional hard constraints such as forcing or avoiding certain entities at chosen locations may simply be added as constraints of the optimization problem. We demonstrated our approach on a variety of real-world datasets.

Future work for our proposed method includes the extension of the plane arrangement data structure to support slanted surfaces and possibly non-planar primitives. Strategies for reducing computational complexity by e.g.\ pruning invalid surface and wall candidates early in the process would improve applicability to larger-scale datasets. Also, connecting our reconstruction methodology with e.g.\ opening and object detection approaches would further enrich the resulting models.

%% file: sec_acknowledgments.tex
\section*{Acknowledgments}

{
\footnotesize
We acknowledge the Visualization and MultiMedia Lab at University of Zurich (UZH) and Claudio Mura for the acquisition of the 3D point clouds, and UZH as well as ETH Zürich for their support to scan the rooms represented in these datasets. Their datasets were used in our evaluation (Figure \ref{fig:synth}).
We also used datasets provided by The Royal Danish Academy of Fine Arts
Schools of Architecture, Design and Conservation (CITA) (Table \ref{tbl:results}), from The ISPRS Benchmark on Indoor Modeling \cite{ISPRS} (Figure \ref{fig:regularization}), and from the Stanford 3D Large-Scale Indoor Spaces Dataset \cite{Buildingparser} (Figure \ref{fig:manual_wall}).
This work was supported by the DFG projects KL 1142/11-1 (DFG Research Unit FOR 2535 Anticipating Human Behavior) and KL 1142/9-2 (DFG Research Unit FOR 1505 Mapping on Demand).
}